\documentstyle[epsfig,aps,multicol]{revtex}

\begin{document}

\title{Permeability and conductivity of platelet-reinforced
membranes and composites}

\author{T. F. Nagy \footnote{ \noindent 
{\bf email:} nagy\_t@pa.msu.edu} and P. M. Duxbury  }

\address{Dept. of Physics/Ast. and 
Center for Fundamental Materials Research,\\
Michigan State University, East Lansing, MI 48824.}

\maketitle

\begin{abstract}
We present large scale simulations of the diffusion constant $D$
of a random composite consisting of aligned platelets
with aspect ratio $a/b>>1$ in a matrix 
(with diffusion constant $D_0$) and find that
$D/D_0 = 1/(1+ c_1 x + c_2 x^2)$, where 
$x= a v_f/b$ and $v_f$ is the platelet volume fraction.  
We demonstrate that for large aspect ratio platelets
the pair term ($x^2$) {\it dominates} suggesting large
property enhancements for these materials.
However a small amount of face-to-face ordering of the platelets
{\it markedly degrades} the efficiency of platelet reinforcement.
\end{abstract}

\begin{multicols}{2}[]
\narrowtext

Thin reinforcing platelets can  be extremely effective at improving
the barrier and in-plane mechanical properties of composites and
membranes.
In particular there has been an explosion of interest in
clay-reinforced thermoplastics, thermosets and rubbers, with target
applications ranging from packaging to cars
\cite{kojima,pinnavaia,giannelis,sherman,garces}. 
To achieve the theoretically promised enhancements requires 
well-aligned and well-dispersed clay platelets in these polymer
matrices.

The traditional theory of composite reinforcement is based on single
inclusion theories which form a basis for self-consistent or
effective-medium approximations.
However the effect of aligned reinforcing platelets is {\it not}
correctly described by single inclusion models
\cite{cussler,fredrickson}, except at very low inclusion
concentrations, even though many publications assume this
approximation \cite{yano,sekelik}. 
The correct variable to use in describing platelet reinforcement
in the large aspect ratio limit is the product of the aspect ratio
($a/b$) {\it times} the volume fraction ($v_f$), $x=a v_f/b$.
Since the aspect ratio of clay platelets ranges from $100-2000$,
$x$ is typically not small as even a 1\% inclusion volume fraction
leads to large values of $x$.
We calculate the diffusion constant in regimes where $x$ is not
small and derive a simple form which represents the data well.  
We find that the {\it quadratic term} dominates (i.e. a term
proportional to $x^2$) and its dominance is due to narrow necks
between platelets which are not included in single inclusion theories.

However, chemically nano-dispersed clays tend to have face-to-face
stacking, which leads to channels of matrix material through which
diffusion is relatively easy.
In a thermodynamic picture this corresponds to a phase separation
of the material into platelet rich and platelet poor regions.
We calculate the dependence of the diffusion constant on face-to-face
alignment and show that the rule of mixtures works effectively.
In clay-polymer materials deleterious platelet-poor channels are
broken up by using extrusion or  mechanical mixing.
To date no-one has found a way to chemically modify the clays so that
they prefer to arrange themselves in a more optimal (e.g. staggered)
array.
Nevertheless, theoretical studies suggest that nematic and staggered
phases can be thermodynamically stable \cite{balazs}.

The leading order term in the reduction in permeability due to platelet
reinforcement is familiar in the composites community, though in a
different context.
There it is well known that a number density of aligned cracks,
$n = N/V$ ($N$ is the number of cracks and $V$ is the sample volume), 
reduces the conductivity of a material of initial conductivity
$\sigma_0$ according to,
\begin{equation}
\sigma = \sigma_0 (1 - \pi a^2 n + \ldots)
\end{equation}
for slits of length $2a$ in a homogeneous two-dimensional medium
\cite{thorpe}, and
\begin{equation}
\sigma = \sigma_0 (1 - \frac{8}{3} a^3 n + \ldots)
\end{equation}
for penny shaped cracks of radius $a$ in a homogeneous three-dimensional
medium\cite{kachanov}.
The perpendicular (ie. normal to the crack surface) conductivity of
cracked solids $\sigma/\sigma_0$, the perpendicular permeability,
$k/k_0$, of platelet-reinforced membranes and the perpendicular
diffusion constant measured in platelet reinforced membranes,
$D/D_0$, are related by, $\sigma/\sigma_0 = k/k_0 = D(1-v_f)/D_0$.
However the experimental results for the permeability of barrier films
are presented as a function of inclusion volume fraction, $v_f$.
This is derived simply from Eq.\ (1) and (2) by using $v_f = n v^*$,
where $v^*$ is the volume of an inclusion.
Thus for barrier membranes, the leading order behavior for aligned
rectangular sticks in two dimensions ($v^* = 4ab$) is, 
\begin{equation}
\frac{k}{k_0} =  1 - \frac{\pi}{4} \frac{av_f}{b} + \ldots,
\end{equation}
and for aligned penny-shaped platelets of radius $a$ and thickness
$2b$ is, 
\begin{equation}
\frac{k}{k_0} =  1 - \frac{4}{3\pi} \frac{av_f}{b} + \ldots
\end{equation}
The importance of the variable $x=a v_f/b$ is evident from these
expresssions.
In the barrier film community the reduction in permeability due to
platelets is approximated in a different way \cite{yano,sekelik}.
There it is argued that for high aspect ratio platelets, the increased
tortuousity of typical diffusion paths $L_p/L_0$ gives the qualitatively
correct reduction in the diffusion constant, i.e. $D/D_0 \sim L_0/L_p$. 
The dependence of path tortuousity on inclusion volume fraction is
given by $ L_p/L_0 = (1 + a v_f/2b)$, where $a/b$ is the platelet aspect
ratio.
Note that this is of the same form as an effective medium theory based
on equations (3) and (4).
However, as we now show this tortuousity argument is {\it qualitatively
incorrect} for platelet reinforced materials and if done correctly leads
to a quadratic (ie. $O((b/(a v_f))^2)$) reduction in the diffusivity
in both two and three dimensions.

We use the resistor respresentation to calculate the effect of tortuous
diffusion paths on the overall conductivity, permeability and diffusivity.
Consider a composite composed of randomly centered, aligned,
non-overlapping, insulating sticks or pennies placed in a matrix of
conductivity $\sigma_0$.
We define the typical perpendicular distance between inclusions to be $l$.
The volume fraction is related to $l$ via, $v_f \approx 2 b/(l+2 b)
\approx 2 b/l$ as the inclusion volume fractions that are observed for
high-aspect ratio materials are typically less than 10\%.
A tortuous path through a random array of these aligned platelets is
approximated by a series combination of resistors each of which has
typical resistance,
\begin{equation}
r_t \approx \frac{\rho_0 a}{l a^{d-2}},
\end{equation}
where $\rho_0=1/\sigma_0$. 
This resistance is calculated by considering a ``neck'' of matrix
material between two adjacent inclusions.
This resistor has typical length of order $a$ (we drop constant prefactors)
and cross-section of order $l a^{d-2}$.
In a film of thickness $t$, the resistance of a tortuous path is then
$R_t = r_t (t/(l+2b)) \approx \rho_0 t a^{3-d}/ l^2$.
In a composite of transverse dimension $L_t$  (perpendicular to the
thickness direction) however, there  are many parallel paths of this
sort and their conductances must be added to approximate the overall
permeability or conductivity.
The typical number of such paths is $(L_t/a)^{d-1}$, so that the typical
resistance of a composite of dimensions $L_t^{d-1} t$ is
$R_f \approx \rho_0 t a^{3-d}/((L_t/a)^{d-1} l^2)$.
We are interested in the conductivity ($\propto$ permeability) which is
related to the resistance by $\sigma = t/R_f (L_t)^{d-1}$.
Using  $v_f = b/l$, we thus find
\begin{equation}
\sigma \approx \sigma_0 (\frac{b}{a v_f})^2 = \sigma_0 \frac{1}{x^2}.
\end{equation}
Note that the result (6) is due to the necks between platelets and
is thus  a pair term and is not included in single inclusion theories.
In order to find typical values for size of this quadratic term, and to
compare its importance with the linear term found from single inclusion 
theories (Eq.\ (3) and (4)), we have carried out large scale simulations.
To compare the above theory with the simulations,  we need to include
both the leading order term (Eq.\ (3) and (4)) and the quadratic term
(6) in our analysis.
The simplest form which contains both of these terms is,
\begin{equation}
\frac{D}{D_0} = \frac{1}{1 + c_1 x + c_2 x^2}
\end{equation}
where $x = a v_f/b$.
As we show below this form works extremely well for aligned
non-overlapping, well-dispersed platelets over a broad range of
concentrations.
Important corrections occur if the platelets stick to each other or if
the platelets overlap (ie. percolation effects) and we shall present the
details of these effects elsewhere. 

Our calculation of the effective diffusivity of clay reinforced membranes
is carried out as follows.
Clay platelets act as effective diffusion barriers to molecules such as
oxygen and water, so we assign them diffusion constant zero.
In contrast many polymers are oxygen and water permeable so we assign the
matrix a finite diffusion constant $D_0$.
We calculate the effective diffusion constant of a polymer containing
volume fraction $v_f$ of aligned platelets, as illustrated in Fig.\ 1.
The effective diffusion constant of these composites is found by
embedding the geometry in large square or cubic lattices.
Random walkers are then started at random locations in these geometries.
The end-to-end distance of these walkers is monitored as a function of time.
Averaging the trajectories leads to the average diffusive behavior
$\langle r^2 \rangle \propto D t $.  
The effective diffusion constant is then found by extracting the slope
of a plot of $\langle r^2 \rangle $ versus $t$.
Our code for the ``blind ant'' \cite{ant} method described above is very
efficient, which enables simulations on  very large lattices over long
times.
This is essential for the large aspect ratio composites discussed here.
Note that $k/k_0 = D(1-v_f)/D_0 $ due to the fact that random walkers
cannot be placed on the zero permeability barriers.
In  experiments, the permeability is measured by placing the barrier in
a pressure gradient, with pressure $p$ of gas (e.g. Oxygen) on one side
of the membrane and the other side maintained at very low pressure.
The steady-state flux of gas, $f$,  through the membrane is measured and
the permeability $k=p/f$.
The diffusion constant is measured by tracking the tracer diffusion of
tagged particles, for example using NMR.

A high precision test of Eq.\ (7) is presented in Fig.\ 2a for aligned
sticks in two dimensions and Fig.\ 2b for aligned squares in three
dimensions.
We plot the quantity $(D_0/D - 1)/x$ which, if Eq.\ (7) is correct,
should be linear, ie. $(D_0/D -1)/x = c_1 +c_2 x$. 
The data of Fig.\ 2 confirms this form remarkably well over the entire
range of inclusion concentrations, even close to the dense packing limit.
A linear fit to the data of Fig.\ 2 yields the coefficients to the
quadratic term,
$c_2^{2d} = 0.165\pm 0.01$ (from Fig.\ 2a), and
$c_2^{3d} = 0.050 \pm 0.005$(from Fig.\ 2b).
The leading order coefficient in the concentration expansion, $c_1$,
is quite poorly converged in two dimensions.
From Fig.\ 2a, it is seen that this $c_1^{2d}$ is still decreasing even
for large aspect ratio platelets.
For the largest aspect ratios we studied its value is 
$c_1^{2d}(a/b=250) = 0.46\pm 0.01$.
In three dimensions the convergence is much better and we find
$c_1^{3d} = 0.44 \pm 0.03$.
The theoretical values (from Eqs.\ (3) and (4)) are
$c_1^{slit} = \pi/4 = 0.785...$, and
$c_1^{penny} = 4/(3\pi) = 0.425...$ respectively.
The numerical simulations are on lattices and so cannot be expected to
be exactly the same as the continuum results (3) and (4).
Nevertheless, it is clear that materials reinforced by well-dispersed 
platelets are {\it not} correctly described by effective medium theory
based on a linear expansion in $x$, or the Nielsen formula
($D/D_0 = 1/(1+x/2)$) which is used in the diffusion community
\cite{yano,sekelik}. 
An important consequence of this result is that the property enhancements
which are theoretically possible from platelet reinforced  materials,
for example clay-polymer nanocomposites, are {\it much larger} than has
previously been suggested or observed.

However the real morphology of, as synthesized, clay-polymer materials
is illustrated in Fig.\ 3 \cite{wang}.
The presence of channels of matrix material (light regions in Fig.\ 3a)
and the strong face-to-face packing of the platelets is evident in Fig\ 3b.
These materials do not provide the barrier performance promised by the
result Eq.\ (7).
The channels of matrix material act as a diffusion ``short-circuit'' so
that the effective diffusion constant remains closer to that of the
matrix material.
In order to make the diffusion paths through the matrix material more
tortuous, materials such as this are mixed in extruders or sheared in
other ways in order to produce better platelet dispersion. 
The effect of various degrees of platelet phase separation on the
diffusion constant is illustrated in Fig.\ 4.
The platelets are initially in a perfectly staggered array and then
one sublattice of the array is shifted by an amount $0\le s \le a$ 
until the platelets are in a perfect face-to-face arrangement.
The rule of mixtures works quite well for these phase separated systems
(see Fig.\ 4), so that,
\begin{equation}
\frac{D(s)}{D_0} = \frac{s}{a}\frac{D(s=1)}{D_0} + 
(1 - \frac{s}{a}) \frac{D(s=0)}{D_0}.
\end{equation}
Clearly even a small amount of phase separation (finite $s$) leads to
a significantly increased diffusion, that is the first term in Eq.\ (8)
dominates for $s> a/(c_2 x^2) = b^2/(c_2 v_f a)$.
In a very real sense, the larger the platelets are the more sensitive
the system is to phase separation.
For example if the reduction in diffusion constant promised by the
formula\ (6) is $1000$ fold, then any matrix channel, which is not highly
tortuous, of size $ 1/1000$ or larger will be deleterious for the
composite performance.
This means that  platelet dispersion must be very good indeed to achieve
the full performance enhancements promised by large aspect ratio platelets. 

In summary, we have shown that the effective diffusion constant (as well
as the conductivity and permeability) of platelet reinforced composites
is not well described by single inclusion theories, even though almost all
experimental  studies in the literature use expressions which are based
on this limit.
This is good news, as it means that the theoretically possible property
enhancements are {\it quadratic} in the volume fraction, rather than linear.
This general result should also apply to many other transport and
mechanical properties of well-dispersed platelet reinforced materials,
and is due to the dominance of narrow necks in large aspect-ratio limit.
However we also showed that face-to-face platelet ordering is extremely
deleterious to the performance of these materials 
(see Fig.\ 4 and Eq.\ (8)), so that achieving the full enhancements
promised by these materials remains a challenging synthesis problem.
 
This work has been supported by the DOE under contract DE-FG02-90ER45418
through a supplement from the CMSN program, by a grant from Ford research
and by the composite materials and structures center at MSU.

%%%%%%%%%%%%%%%%%%%%%%%%%%%%%%%%%%%%%%%%%%%%%%%%%%%%%%%%%%%%%%%%%%%%%%%%%%%
% REFERENCES
%%%%%%%%%%%%%%%%%%%%%%%%%%%%%%%%%%%%%%%%%%%%%%%%%%%%%%%%%%%%%%%%%%%%%%%%%%%

\begin{figure}
\centerline{\epsfig{file=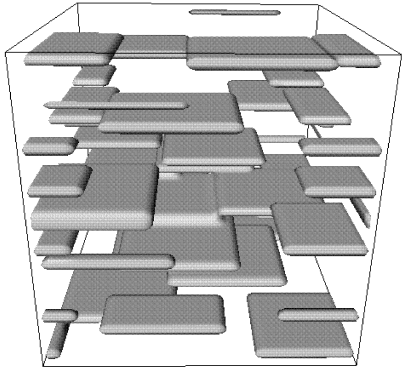,width=7cm,angle=0}}
\vspace{0.2in}
\caption{Aligned square platelets randomly embedded in a three-dimensional 
cube. The aspect ratio $a/b=25$, the boundary conditions are periodic in
all directions.}
\label{fig1}
\end{figure}

\begin{figure}
\centerline{\epsfig{file=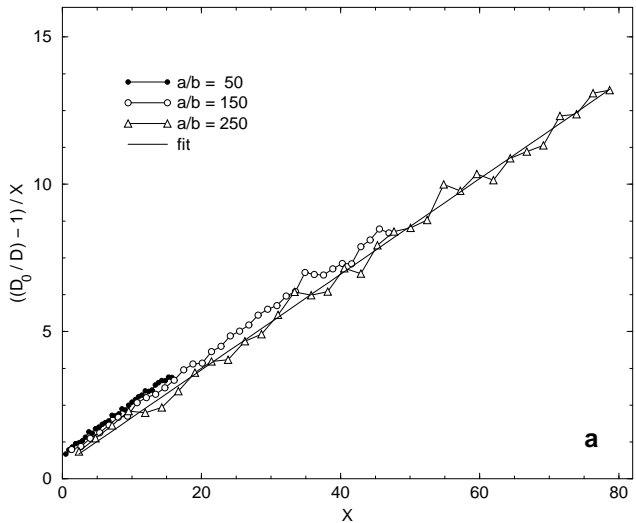,width=7cm,angle=270}}
\vspace{0.2in}
\centerline{\epsfig{file=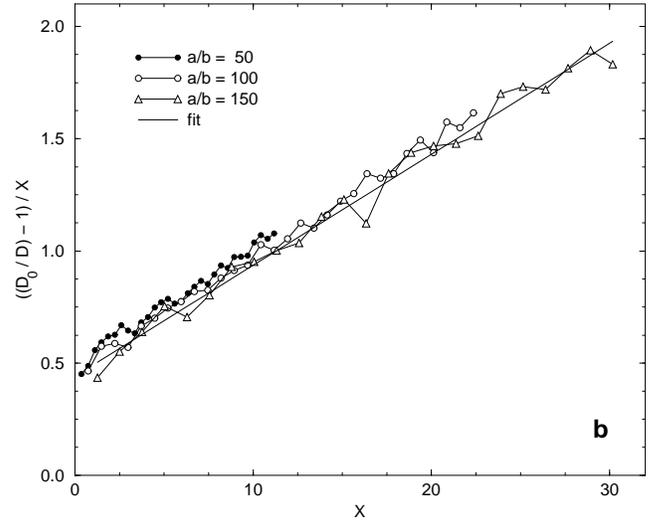,width=7cm,angle=270}}
\vspace{0.2in}
\caption{Tests of the form (7) of the text.  A plot of $(D_0/D - 1)/x$
as a function of $x = a v_f/b$, where $a/b$ is the platelet aspect ratio
and $v_f$ their volume fraction.
The solid lines are fits to the largest aspect ratio data. 
{\bf (a)} Data for aligned slits randomly placed, without overlap, onto
a square lattice.
The simulations were carried out on square lattices of size $2048^2$,
over $6 \times 10^4$ steps in the blind ant algorithm (see text) and
averaged over $30,000$ configurations.
{\bf (b)} Data for aligned  squares (see Fig.\ 1) placed, without
overlap, onto a cubic lattice.  
The simulations were carried out on cubic lattices of size $512^3$, over
$5 \times 10^4$ steps in the blind ant algorithm (see text) and averaged
over $20,000$ configurations.} 
\label{fig2}
\end{figure}

\begin{figure}
\centerline{\epsfig{file=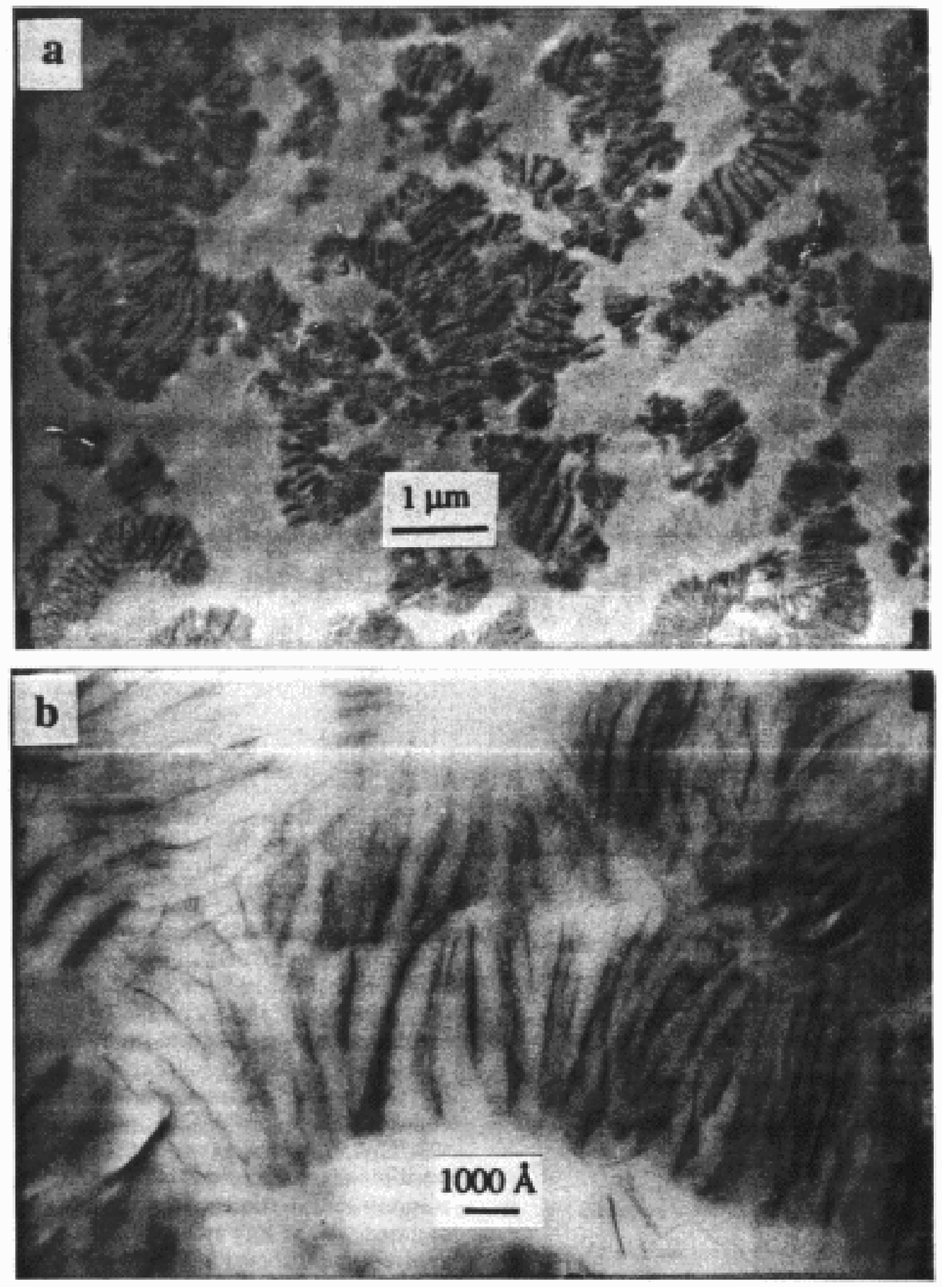,width=7cm,angle=0}}
\vspace{0.2in}
\caption{Experimental platelet morphology at two length scales.
{\bf (a)} The upper figure shows that there is a phase separation
into platelet rich and platelet poor regions.
{\bf (b)} The lower figure illustrates the face-to-face alignment
of the platelets.
The dark regions are clay, the light regions are the polymer
(from ref.\ [14]).}
\label{fig3}
\end{figure}

\begin{figure}
\centerline{\epsfig{file=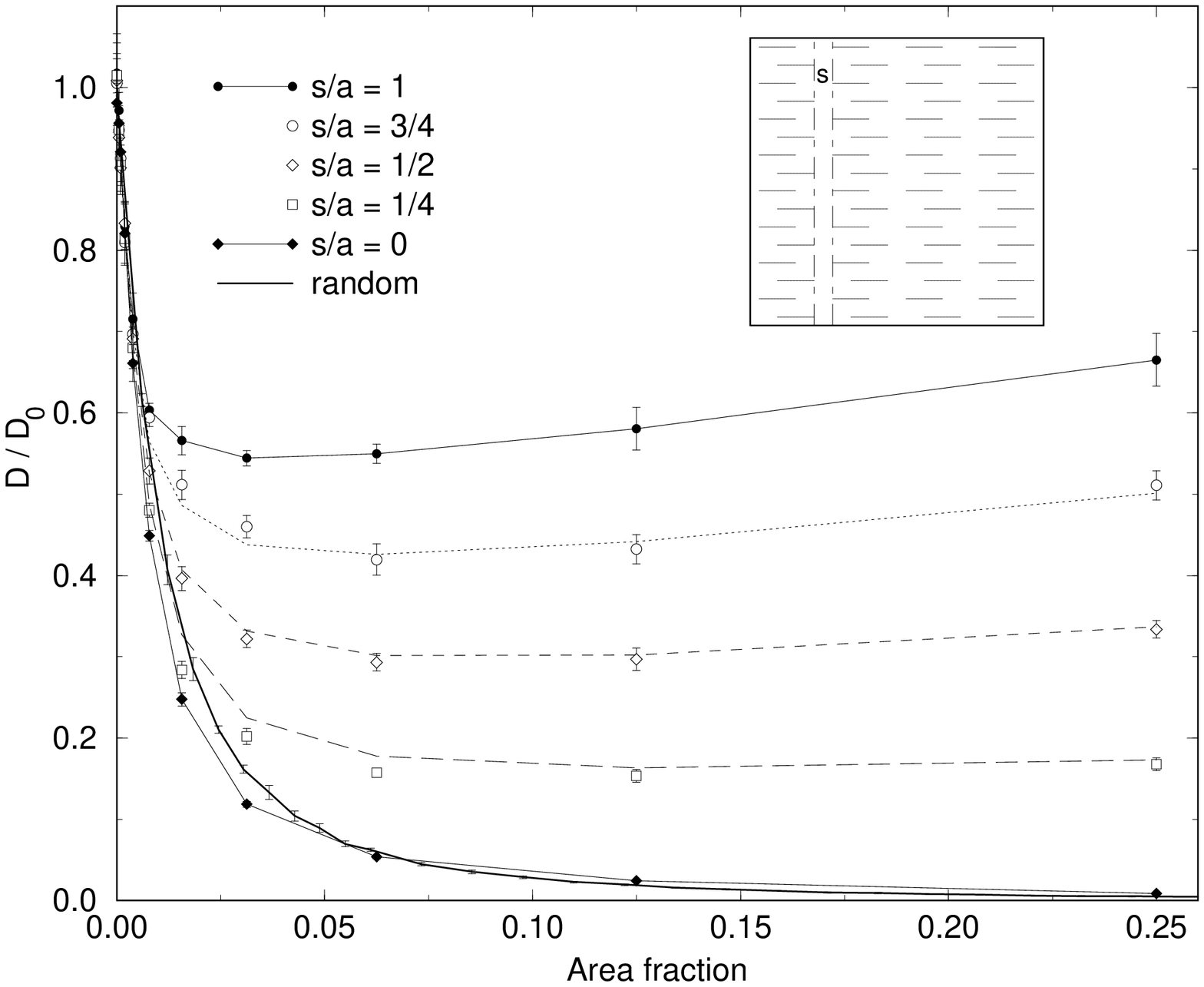,width=7cm,angle=270}}
\vspace{0.2in}
\caption{The effective diffusion constant as a function of face-to-face
alignment of the platelets (of aspect ratio $a/b=128$).
The inset shows the geometry that was used for the calculations.
Also included is the data for a random system with the same aspect ratio
and it shows that the random system is similar to the perfectly
staggered array of barriers, at least at the resolution of this plot.
The three plots at intermediate $s$ (dots, small dashes and large dashes)
are fits using the rule of mixtures (Eq.\ (8)).}
\label{fig4}
\end{figure}

\end{multicols}

\begin{references}
\bibitem{kojima}
Y. Kojima, A. Usuki, M. Kawasumi, A. Okada, T. Kurauchi
and O. Kamigaito, J. Poly. Sci. {\bf A31}, 983 (1993).
\bibitem{pinnavaia}
P. Lan, P.D. Kaviratna and T.J. Pinnavaia, 
J. Chem. Mater. {\bf 7}, 2144 (1995).
\bibitem{giannelis}
E.P. Giannelis, R. Krishnamoorti and E. Manias, 
Adv. Poly. Sci. {\bf 138}, 107 (1999).
\bibitem{sherman}
L.M. Sherman, Plastics Technology, June issue, 52 (1999).
\bibitem{garces}
J.M. Garces, D.J. Moll, J. Bicerano, R. Fibiger and D.G. McLeod,
Adv. Mat. {\bf 12}, 1835 (2000).
\bibitem{cussler}
E.L. Cussler, S.E. Hughes, W.J. Ward III and R. Aris,
Journal of Membrane Science {\bf 38}, 161 (1988); 
W.R. Falla, M. Mulski and E.L. Cussler, 
Journal of Membrane Science {\bf 119},
129 (1996).
\bibitem{fredrickson}
G.H. Fredrickson and J. Bicerano, J. Chem. Phys. 
{\bf 110}, 2181 (1999).
\bibitem{yano}
K. Yano, A. Usuki, A. Okada, T. Kurauchi and O. Kamigaito,
J. Poly. Sci. {bf A31}, 2493 (1993).
\bibitem{sekelik}
L.E. Nielsen, J. Macromol. Sci. (Chem.) {\bf A1}, 929 (1967);
D.J. Sekelik, E.V. Stepanov, S. Nazarenko, D. Schiraldi, A. Hiltner and
E. Baer, J. Poly. Sci. {\bf B37}, 847 (1999).
\bibitem{balazs}
V.V. Ginzburg, C. Singh, A.C. Balazs,
Macromolecules {\bf 33}, 1089 (2000).
\bibitem{thorpe}
For a recent discussion see M.F. Thorpe, Proc. Roy. Soc. {\bf A437},
215 (1992) - Eq. (43).
\bibitem{kachanov}
For a recent discussion see B. Shafiro and M. Kachanov, J. Appl. Phys.
{\bf 87}, 8561 (2000) - Eq. (9).
\bibitem{ant}
L. Puech and R. Rammal, J. Phys. {\bf C35}, L1197 (1983); J. Tobochnik,
D. Laing, G. Wilson, Phys. Rev. {\bf A41}, 3052 (1990); 
I.C. Kim and S. Torquato, J. Appl. Phys. {\bf 74},
1844 (1993).
\bibitem{wang}
M.S. Wang and T.J. Pinnavaia, Chemistry of Materials {\bf 6}, 468 (1994).
\end{references}
\end{document}